\documentclass[pre, twocolumn, groupedaddress, superscriptaddress, amsmath, amssymb, floatfix,showpacs]{revtex4-1}

\usepackage{graphicx}
\usepackage{dcolumn}
\usepackage{xcolor}
\usepackage{textcomp,gensymb}
\usepackage{bm}
\makeatletter
\renewcommand{\@fnsymbol}[1]{
    \ifcase#1 \or *\or §\or  ¶\or §\or ¶\or **\or \else \mathsection\fi
}

\begin{document}

\title{The Spontaneous Cascading Mechanism Behind Critical Phenomena in Self-Coupled Lasers
}

\author{Jiaoqing Wang}
\thanks{J.W. \& Y.K.C contributed equally to this work.}
\affiliation{
Department of Physics, The Jack and Pearl Resnick Institute for Advanced Technology, Bar-Ilan University, Ramat-Gan, 5290002 Israel}

\author{Yael Kfir-Cohen}
\thanks{J.W. \& Y.K.C contributed equally to this work.}
\affiliation{
Department of Physics, The Jack and Pearl Resnick Institute for Advanced Technology, Bar-Ilan University, Ramat-Gan, 5290002 Israel}

\author{Chenni Xu}
\affiliation{
Department of Physics, The Jack and Pearl Resnick Institute for Advanced Technology, Bar-Ilan University, Ramat-Gan, 5290002 Israel}

\author{Bnaya Gross}
\affiliation{
Network Science Institute, Northeastern University, Boston, MA 02115}
\affiliation{
Department of Physics, Northeastern University, Boston, MA 02115}

\author{Aswathy Sundaresan}
\affiliation{
Department of Physics, The Jack and Pearl Resnick Institute for Advanced Technology, Bar-Ilan University, Ramat-Gan, 5290002 Israel}

\author{Shlomo Havlin}
\affiliation{
Department of Physics, The Jack and Pearl Resnick Institute for Advanced Technology, Bar-Ilan University, Ramat-Gan, 5290002 Israel}

\author{Patrick Sebbah}
\thanks{Corresponding author: patrick.sebbah@biu.ac.il}
\affiliation{
Department of Physics, The Jack and Pearl Resnick Institute for Advanced Technology, Bar-Ilan University, Ramat-Gan, 5290002 Israel}

\date{\today}

\begin{abstract}
The basic physics of lasers is characterized by a second-order continuous phase transition at the critical lasing threshold. Nevertheless, laser bistability with abrupt transitions has been reported in some laser systems, but its underlying mechanism has never been explored. Here we study experimentally and theoretically a novel nonlinearly self-coupled laser system. We show both experimentally and theoretically that this system experiences spontaneous cascading that yields an abrupt mixed-order transition. At the critical point, a long-lived cascading plateau is observed, characterized by a critical branching factor equal to one. When deviating from criticality, the branching factor departs monotonically from one. The critical scaling close to and at the critical point resembles similar phenomena observed recently in other interdependent systems, suggesting a common universal cascading origin for abrupt transitions. Our results shed light on the cascading mechanism of abrupt transitions in laser systems, which can be utilized for future research and applications.

\end{abstract}
\maketitle
\def\thefootnote{\star}\footnotetext{These authors contributed equally to this work.}

Since the early days of laser research, phase transitions and critical phenomena in lasers have attracted significant attention. It has been recognized that the transition above threshold in lasers exhibits characteristics akin to a non-equilibrium phase transition. 
In most laser systems, the transition between spontaneous emission and coherent stimulated emission occurs smoothly, resembling the continuous state change observed in a second-order phase transition \cite{Scott1975,DeGiorgio1970,Graham1970LaserlightF,Grossmann1971,DOHM19721273}.
However, under specific configurations, abrupt and discontinuous changes in laser intensity have been observed, indicating a first-order abrupt phase transition. This abrupt transition type was first reported in gas lasers \cite{PismaZhETF.7.3} and semiconductor lasers \cite{LASHER1964707, Nathan1965} by incorporating elements such as saturable absorbers into the laser cavity. 
Moreover, a hysteresis effect observed in the laser output intensity versus pump input characteristics was interpreted as analogous to a liquid-gas transition near the critical point \cite{kazantsev1970quantum}.
Since then, the concept of bistability at the origin of the hysteresis loop observed in lasers has garnered significant attention in developing bistable optical devices, which are crucial for advancing all-optical computing \cite{article}, in semiconductor lasers \cite{Nathan1965,Kawaguchi1982,1073533,harari2018topological} and in intensity-coupled lasers \cite{Dutta1984,Oudar1984}.

Rosen \textit{et al}.~\cite{Rosen2010} further investigated abrupt transitions in lasers by measuring critical exponents near the transition. These abrupt transitions provide a theoretical basis for understanding laser instabilities \cite{Krger2015} and optical turbulence \cite{PhysRevLett.45.709}, and opened new applications in chaos communications, random number generation, and reservoir computing \cite{Sciamanna2015,Ohtsubo2017}. 
Despite these findings, the underlying mechanisms driving such abrupt transitions in laser systems remain unclear. Understanding the conditions and the mechanisms that govern these sharp changes could have significant implications for understanding phase transitions and, particularly, designing more efficient and tunable lasers with precise control over their emission properties.
\par
In this Letter, we investigate both theoretically and experimentally a self-coupled laser, a system that incorporates a computer-controlled intensity feedback on the pump. By selecting a nonlinear coupling function, we demonstrate that it can exhibit an abrupt transition. Our investigation reveals a critical slowing down of the laser's transient time as the system approaches the transition point, characterized by critical exponents. This observation is explained here in terms of a spontaneous chain reaction of critical cascade changes within the system. The feedback mechanism operates slowly at criticality with a critical branching factor equal exactly to one, ultimately leading to a cascade of small changes that yield a sudden, abrupt laser switch-on as well as switch-off. This is similar to the cascading phenomena observed in the extensively studied framework of interdependent networks theory \cite{buldyrev-nature2010,dong-pre2014} that has recently been identified in several physical systems, including interdependent superconductors \cite{Bonamassa2023, article_Communications} and interdependent ferromagnetic networks \cite{gross2024microscopic}. In these systems, dependencies between nodes in different networks allow failure propagation, eventually leading to a cascading failure and mixed-order transition with a branching factor of one at criticality. Our experimental and theoretical analysis of the underlying mechanism of the self-coupled laser system includes measurements of the critical exponents, which are found to be identical to those observed in interdependent networks. This result proves that self-coupled lasers belong to the mixed-order universality class of interdependent systems \cite{bonamassa2025hybrid}. Furthermore, this shows that the universality class of interdependent systems is broader and is determined by the underlying cascade mechanism of the system and not the network structure, which is absent in self-coupled lasers.

\begin{figure*}[ht]
		\centering
{\includegraphics[width=\linewidth]{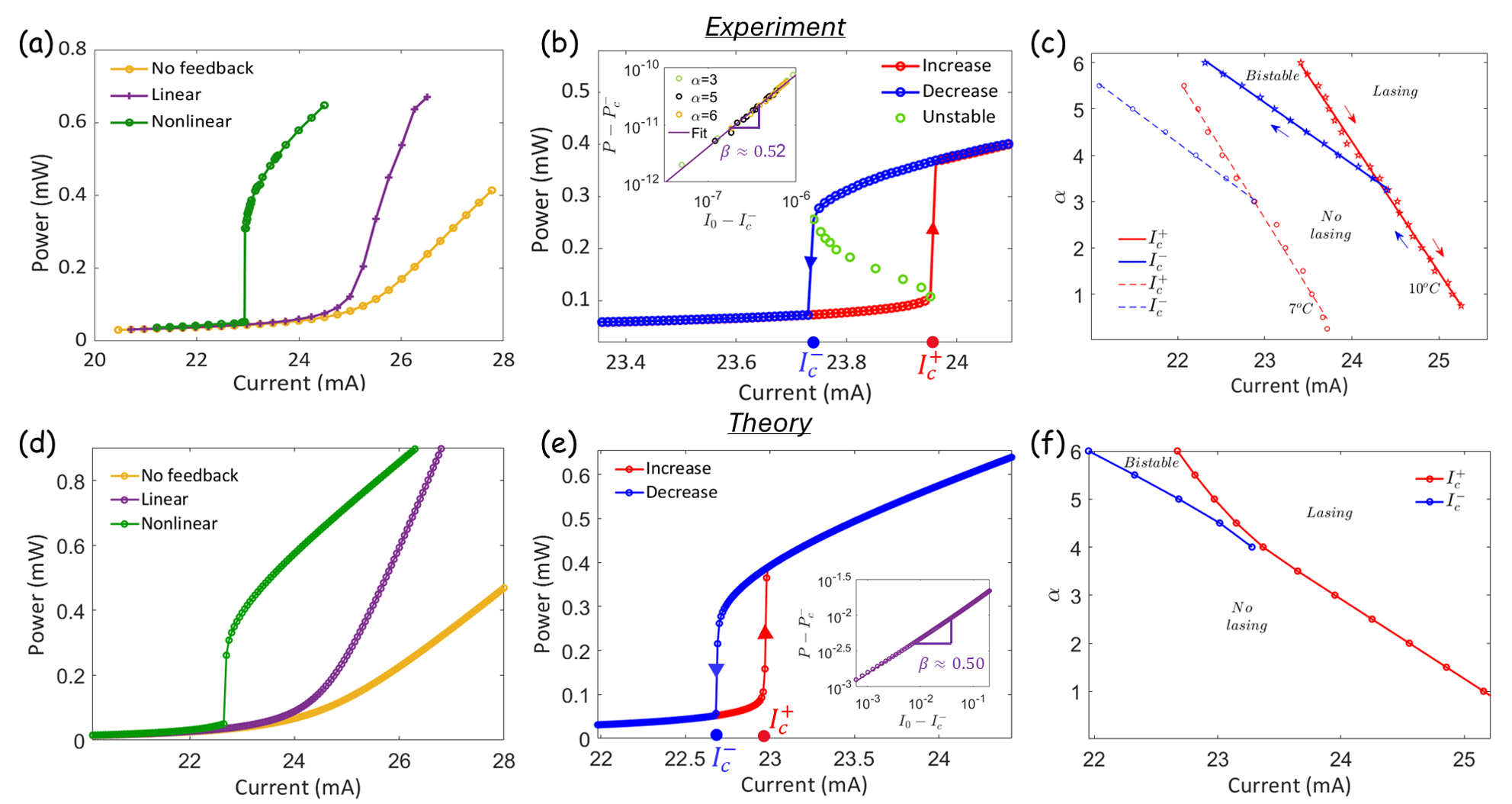}}
    \caption{
        \textbf{Self-coupled laser.} \textbf{(a)} Experimental and \textbf{(d)} theoretical plot of laser output power vs. pump current, for the laser (yellow) without feedback, (purple) with linear feedback, and (green)  with nonlinear feedback. Measurements were made at 10 $^\circ\mathrm{C}$, with coupling strength $\alpha=5$. 
        \textbf{(b)} Measurement and \textbf{(e)} numerical calculation of the hysteresis loop $P(I)$ for the nonlinear coupling case. The unstable solution branch is shown in green open circles. Insets: Logarithmic plot of $P-P_c^-$ near the transition $I_c^-$, for three values of the coupling strength, $\alpha=3, 5, 6$.
        \textbf{(c)} Experimental phase diagram ($I$,$\alpha$) obtained for 7 $^\circ\mathrm{C}$ (light-colored symbols) and 10 $^\circ\mathrm{C}$ (dark-colored symbols). \textbf{(f)} Theoretical phase diagram obtained from Eqs.~\eqref{eq:classic_laser} and~\eqref{eq:Rp/dt}.
        }
    \label{fig:figure experiment1}
\end{figure*}
\par
 \textbf{Self-coupled laser - experiments -} The experimental setup consists of a temperature-controlled, electrically pumped commercial laser diode (For more details see \cite{SM}). Self-coupling is implemented by dynamically adjusting the injection current based on the measured laser output power. More specifically, a computer processes the optical signal and updates the injection current according to a predefined feedback function $f$, such that
\[
I_{i+1} = I_0 + f(P_i),
\]
where $P_i$ is the laser output power measured at injection current $I_i$, $I_0$ is the initially set current, and $i$ denotes the iteration step. The system is considered to have reached steady state when the output power converges within a specified tolerance, i.e.,
\[
|P_{i+1} - P_i| < \epsilon_P \quad,
\]
where \( \epsilon_P =10^{-4}\)mW is an arbitrary convergence threshold.
Compared to all other laser systems with optoelectronic feedback described in the literature \cite{lin2003nonlinear,abarbanel2001synchronization,ghalib2012quantum,satyan2009precise}, the advantage of this method is the unique ability to choose the function $f(P)$ that couples the photodetected laser output $P$ with the injection current $I_{0}$.

When nonlinear coupling is introduced, an abrupt transition occurs, as illustrated in Fig.~\ref{fig:figure experiment1}(a). This behavior contrasts with the shown case of linear coupling, $f(P)=\alpha P$, where the laser transition remains continuous and manifests only as an increase in slope efficiency with increasing $\alpha$. With nonlinear coupling, the laser turns on abruptly at the critical input intensity $I_{c}^{+}$ and switches off abruptly at $I_{c}^{-}$, forming a hysteresis loop as shown in Fig.~\ref{fig:figure experiment1}(b). This loop can be accurately traced experimentally by increasing/decreasing the injection current $I_0$ in small increments of 10$\mu$A. The unstable solution branch shown in Fig.~\ref{fig:figure experiment1}(b) is reconstructed using the method detailed in \cite{SM}.

The broadness of the hysteresis, $I_{c}^{+}-I_{c}^{-}$ is found to depend on the coupling strength $\alpha$. This leads to the phase diagram ($I$,$\alpha$) presented in Fig.~\ref{fig:figure experiment1}(c) for two values of temperature, $T$=10\degree C and $T$=7\degree C. In this phase diagram, the critical currents, $I_c^-(\alpha)$ and $I_c^+(\alpha)$, appear as straight lines separating the lasing and non-lasing phases. Bistability starts at the tricritical point for which $\alpha=\alpha_{c}$. For $\alpha < \alpha_c$, $I_{c}^{-} = I_{c}^{+}$ and the transition is continuous, while for $\alpha > \alpha_c$ the transition is abrupt with bistability characterized by $I_{c}^{+} > I_{c}^{-}$. 
\par
\textbf{Critical behavior - experiments -} Mixed-order phase transitions are a unique class of phase transitions, displaying both an abrupt change similar to first-order transitions and scaling laws and critical exponents near the critical point similar to continuous second-order transitions \cite{mukamel2024mixed,boccaletti2016explosive,gross2022fractal,alert2017mixed,korbel2025microscopic}. To classify the nature of the transition and its universality class, we examine the critical behavior near the critical points. We find that the laser power near both critical points $I_{c}^{-}$ and $I_{c}^{+}$ follows a power-law:
\begin{equation}
\label{eq:beta}
\begin{gathered}
P - P_{c}\sim (I-I_{c})^{\beta}
\end{gathered}
\end{equation}
where the exponent $\beta= 0.52$ near $I_c^-$  (see inset of Fig.~\ref{fig:figure experiment1}(b)) and $\beta= 0.50$ near $I_c^+$ \cite{SM}, thus confirming the expected square-root scaling predicted for interdependent networks \cite{buldyrev-nature2010}, and similar to the results of Rosen \textit{et al}.~\cite{Rosen2010}. 
Interestingly, the inset of Fig.~\ref{fig:figure experiment1}(b) shows that logarithmic plots for different values of $\alpha$ at fixed temperature collapse onto a single line, indicating that the proportionality coefficient in Eq.~\eqref{eq:beta} is independent of $\alpha$.
\par
\begin{figure*}[ht]
		\centering
{\includegraphics[width=\linewidth]{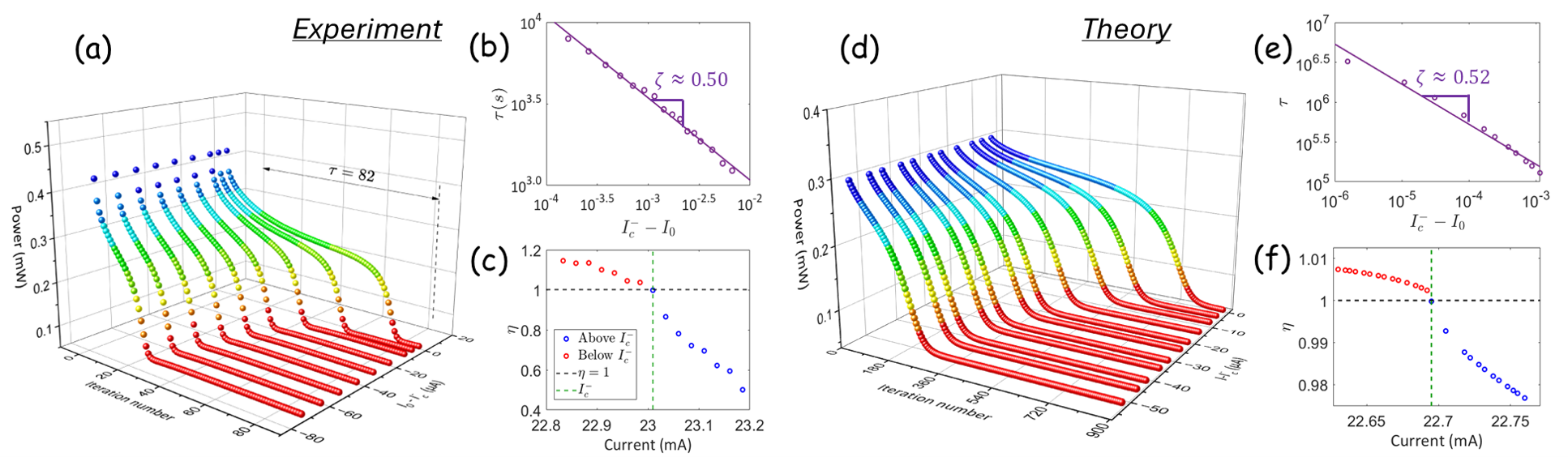}}
    \caption{
        \textbf{Plateau and branching factor.}
        \textbf{(a,d)} Optical power vs. time (iteration number) in response to a short perturbation of the injection current just below the critical point $I_c^-$, for decreasing difference $I_0-I_c^-$. (a) Experiment; (d) Numerical calculations. \textbf{(b,e)} Logarithmic plot of the time duration of the plateau, $\tau$, vs. $I_0-I_c^-$ and the critical exponent $\zeta$. (b) experiment; (e) Numerical calculations. \textbf{(c,f)} Branching factor $\eta$. (c) Experiment; (f) Numerical calculations. All measurements were made at 10 $^\circ\mathrm{C}$ for $\alpha=5$.
    }
    \label{fig:figure 3}
\end{figure*}

\textbf{Plateau - experiments -} Critical exponents are universal descriptors that tell us how the system behaves near the transition, but not why and how the system transitions. To elucidate the mechanism underlying this self-coupled laser mixed-order transition, we investigate the dynamics of the laser in the vicinity of the critical point. To this end, we prepare the system in a non-equilibrium state and monitor its relaxation toward equilibrium. We first consider the non-lasing regime, $I_0 < I_{c}^{-}$, and apply a brief perturbation $\delta I$ such that $I_0+\delta I > I_{c}^{-}$, momentarily driving the system into the lasing state. The optical output power $P$ is monitored and the transient injection current $I=I_0+f(P)$ is iteratively updated until the system returns to its non-lasing steady state (Fig.~\ref{fig:figure 3}(a)). As $I_0$ approaches the lower  critical threshold, $I_{c}^{-}$,  the relaxation exhibits a pronounced slowdown, forming a characteristic  \textit{plateau}. The duration $\tau$ of this plateau is measured and found to follow a power law scaling near the transition:
\begin{equation}
\label{eq:tau}
\begin{gathered}
\tau \propto(I-I_{c}^{-})^{-\zeta}, 
\end{gathered}
\end{equation}
with a critical exponent $\zeta$ = 0.50 (see Fig.~\ref{fig:figure 3}(b)). A similar critical slowing down is observed when probing the system from above the upper threshold \( I_c^{+} \) \cite{SM}. In this case, a brief negative perturbation \( I_0 - \delta I \) temporarily turns the laser off. As \( I_0 \) approaches \( I_c^{+} \), the relaxation back to the lasing state develops a plateau whose duration scales as a power law with the same exponent \( \zeta \).
\par
\textbf{Branching factor - experiments -} The emergence of the plateau, along with its scaling properties, signals a progressive transformation that governs the slowing down of the laser dynamics near the critical point. The quantity that best probes this cascading mechanism is the branching factor, $\eta$. The branching factor measures the average number of secondary events triggered during the cascade by a single primary event. This control parameter has been used to model e.g. epidemic or forest fire spread, information cascades, neural avalanches, to determine whether an event dies out or leads to a large cascade.  When $\eta$ reaches unity, the system undergoes a phase transition.
In the coupled-laser system, the branching factor can be obtained by measuring $\eta =\langle\frac{P_{i+1}-P_{i}}{P_{i}-P_{i-1}}\rangle$, where $P_i$ is the transient laser output power at step $i$ and $\langle...\rangle$ stands for time averaging over the plateau duration. 
As the initial injection current, $I_0$, is reduced stepwise toward the transition $I_c^-$, the branching factor $\eta$ is found to increase from below one (lasing state) to one, at the critical point, when the laser stops emitting. Beyond that point, the branching factor becomes larger than one, as illustrated in Fig.~\ref{fig:figure 3}(c). Measurement methods are detailed in \cite{SM}.
\par

\textbf{Self-coupled laser - theory -} To elucidate the mechanisms underlying the above experimentally observed abrupt transitions and cascading phenomena, we develop a theoretical framework based on a simplified analytical model of the self-coupled laser. This model describes a single-mode laser in terms of three key quantities: the number of photons \( n \), the number of excited atoms \( N \), and the injection current \( I \). As will be shown, this model successfully captures the essential features of the experimentally observed phenomena. \\
In the absence of coupling, their time evolution is governed by the standard set of coupled rate equations for semiconductor lasers \cite{Agrawal2002}:
\begin{equation}
\label{eq:classic_laser}
\begin{gathered}
\frac{dn}{dt}=KnN-\gamma_{c}n+b\gamma_{2}N\\
\frac{dN}{dt}=\frac{I}{q}-KnN-\gamma_{2}N .
\end{gathered}
\end{equation}
Here, $\gamma_c$ is the photon loss rate, $\gamma_2$ the atomic decay rate, $K$ the stimulated-transition coefficient, $q$ the carrier charge, and $b$ the spontaneous-emission factor. Numerical results use $\gamma_{c}=7.479\,\mathrm{s}^{-1}$, $\gamma_{2}=4.291\,\mathrm{s}^{-1}$, $q=1$, $K=1.5$, and $b=10^{-3}$, while the analytical treatment assumes $b \approx 0$. \\
An additional rate equation
\begin{equation}
\label{eq:Rp/dt}
\begin{gathered}
\frac{dI}{dt}=I_0-I+f(n)
\end{gathered}
\end{equation}
captures the feedback coupling between $I$ and $n$, where $I_0$ denotes the initial current injected at $t = 0$.
In the steady state, the pump current becomes $I = I_0+f(n)$, reflecting the feedback from the laser output, given the proportionality between the photon density $n$ and the power $P$.
\par
We begin by introducing the linear coupling function, $f(n)=\alpha n$. Below the critical threshold, the trivial solution $n = 0$ is the stable solution, while above the threshold, a new non-zero steady-state solution emerges: $n=\frac{I-\gamma_{c}\gamma_{2}/K}{\gamma_{c}-\alpha}$.
Despite the coupling, the transition remains continuous but exhibits a steeper linear slope of $1/(\gamma_{c}-\alpha)$, compared to $1/\gamma_{c}$ in the absence of coupling ($\alpha=0$), as illustrated in Fig.~\ref{fig:figure experiment1}(d).

Next, we examine the system under a nonlinear coupling of the form, $f(n)=\alpha \cdot \tanh(qn)$.The coupled equations \ref{eq:classic_laser} and \ref{eq:Rp/dt} are integrated numerically using the Euler method until transients vanish, and the steady state is recorded.
This time, the self-coupled laser exhibits an abrupt jump (Fig.~\ref{fig:figure experiment1}(d)), closely reproducing the experimental observations (Fig.~\ref{fig:figure experiment1}(a)). 
The steady-state solution of the nonlinearly-coupled laser satisfies the following equation:
\begin{equation}
\label{eq:tanh}
\begin{gathered}
\gamma_{c}\cdot n=I_0-\frac{\gamma_{c}\gamma_{2}}{K}+\alpha \cdot \tanh(qn)
\end{gathered}
\end{equation}
The multiplicity of solutions found - 2 stable and 1 unstable - indicates the presence of a bistable regime, as confirmed by numerically solving the laser equations (Fig.~\ref{fig:figure experiment1}(e)). The resulting hysteresis loop between critical points, $I_c^-$ and $I_c^+$, is characterized by the coexistence of two stable states: an active (lasing) state and an inactive (non-lasing) state. 
By systematically varying the parameters $\alpha$ and $I$, we qualitatively reproduce the experimental phase diagram, as shown in Fig.~\ref{fig:figure experiment1}(f). \\
\textbf{Critical behavior - theory -} To further validate our simplified theoretical model, we now examine the critical exponents and scaling behavior near the abrupt transition. 
The critical exponent $\beta$, characterizing the growth of the optical power above the critical injection current $I_c^-$, is extracted by numerically solving Eq.~\eqref{eq:tanh}.
The best fit of $P-P_c$ versus $I_0-I_c^-$ yields a slope of $0.50$ as shown in the inset of Fig.~\ref{fig:figure experiment1}(e), in excellent agreement with the experimentally determined value $\beta =0.52$.
This result is further supported by an independent analytical derivation, which yields $\beta =0.5$ and confirms the independence with $\alpha$ of the proportionality coefficient \cite{SM}.

\textbf{Plateau - theory -} Next, the dynamics of the self-coupled laser are investigated theoretically by solving the coupled equations 
Eqs.~\eqref{eq:classic_laser}~and~\eqref{eq:Rp/dt}. The system is perturbed below $I_{c}^-$ and the time evolution of the laser output returning to the steady state is computed. As the initial condition approaches the critical point, a pronounced critical slowdown is observed, characterized by the emergence of a plateau of length $\tau$. Figure~\ref{fig:figure 3}(e) shows $\tau$ as a function of $I_{c}^{-}-I$ on a double logarithmic scale, confirming a power-law scaling with an exponent $\zeta = 0.52$, in close agreement with the experimentally measured value.

\textbf{Branching factor - theory -} The branching factor is defined as the average number of photons generated per photon from the previous time step. We find that $\eta > 1$ above the critical injection current and $\eta < 1$ below threshold, while at the critical point $\eta \to 1$ from both directions (Fig.~\ref{fig:figure 3}(f)), in agreement with the experiment. At this stage of the transition, each photon reproduces, on average, exactly one photon---revealing the hallmark of a self-sustained critical process. The theory thus uncovers, at the microscopic scale, the cascading dynamics that manifest experimentally as the abrupt phase transition.

\textbf{Discussion -} In summary, we identify a universal microscopic cascading mechanism that drives the macroscopic abrupt phase transition observed experimentally in a nonlinearly self-coupled laser. This transition is marked by a long-lived transient plateau near criticality and a branching factor converging to unity, and is characterized by novel critical exponents that capture the scaling of the dynamics. Our findings place this coupled laser system within the broader class of physical systems exhibiting mixed-order transitions---featuring both discontinuities and critical behavior \cite{bonamassa2025hybrid,mukamel2024mixed,boccaletti2016explosive,gross2022fractal,alert2017mixed,korbel2025microscopic}---akin to those identified in the theory of interdependent networks \cite{buldyrev-nature2010,dong-pre2014}. Furthermore, these results demonstrate that the universality class of interdependent systems is broader than previously thought, being dictated by the underlying cascade mechanism rather than by network structure, which is entirely absent in self-coupled lasers.

In \cite{SM}, we explore an alternative nonlinear coupling function that gives rise to double bistability, further enriching the pathways for controlling phase transitions in self-coupled systems. Together, our results not only open new avenues for fundamental research into critical phase transitions and cascading phenomena, but also point toward practical applications in the design of optical switches, photonic memory elements, and dynamically reconfigurable laser systems.\\

\textbf{Acknowledgments -} S.H. and P.S. thank Prof. M. Segev for initiating this collaboration. P.S. acknowledges the precious help of his lab manager, Leonid Wolfson. This work was supported by The Israel Science Foundation grants 3777/25, 1698/22, 2630/20, 189/19; the United States–Israel Binational Science Foundation NSF/BSF Grants 2015694, 2019740 and 2020245, the BSF Grant 2022158; the Israel Ministry of Innovation, Science $\&$ Technology, Grant No. 01017980; The Science Minister-Smart Mobility, Grant No. 1001706769; EU H2020 Project OMINO, Grant No. 101086321 and the Israeli VATAT project on power-grid. B.G. acknowledges the support of the Fulbright Postdoctoral Fellowship Program. C.X. acknowledges the Excellence Fellowship for international postdoctoral researchers funded by the Israel Academy of Sciences and Humanities and the Council for Higher Education of Israel. J.W. has been supported by the Excel@BINA scholarship from the Nanocenter at Bar Ilan University.

\bibliographystyle{apsrev4-2}
\bibliography{references}

\end{document}